\documentclass[twoside,twocolumn,9pt]{article}
\usepackage{extsizes}
\usepackage[super,sort&compress,comma]{natbib} 
\usepackage[version=3]{mhchem}
\usepackage[left=1.5cm, right=1.5cm, top=1.785cm, bottom=2.0cm]{geometry}
\usepackage{balance}
\usepackage{mathptmx}
\usepackage{sectsty}
\usepackage{graphicx} 
\usepackage{lastpage}
\usepackage[format=plain,justification=justified,singlelinecheck=false,font={stretch=1.125,small},labelfont=bf,labelsep=space]{caption}
\usepackage{float}
\usepackage{fancyhdr}
\usepackage{fnpos}
\usepackage[english]{babel}
\addto{\captionsenglish}{%
  
}
\usepackage{array}
\usepackage{droidsans}
\usepackage{charter}
\usepackage[T1]{fontenc}
\usepackage[usenames,dvipsnames]{xcolor}
\usepackage{setspace}
\usepackage[compact]{titlesec}
\usepackage{hyperref}

\usepackage{epstopdf}

\definecolor{cream}{RGB}{222,217,201}

\begin{document}

\pagestyle{fancy}
\thispagestyle{plain}
\fancypagestyle{plain}{
\renewcommand{\headrulewidth}{0pt}
}

\makeFNbottom
\makeatletter
\renewcommand\LARGE{\@setfontsize\LARGE{15pt}{17}}
\renewcommand\Large{\@setfontsize\Large{12pt}{14}}
\renewcommand\large{\@setfontsize\large{10pt}{12}}
\renewcommand\footnotesize{\@setfontsize\footnotesize{7pt}{10}}
\makeatother

\renewcommand{\thefootnote}{\fnsymbol{footnote}}
\renewcommand\footnoterule{\vspace*{1pt}%
\color{cream}\hrule width 3.5in height 0.4pt \color{black}\vspace*{5pt}} 
\setcounter{secnumdepth}{5}

\makeatletter 
\renewcommand\@biblabel[1]{#1}            
\renewcommand\@makefntext[1]%
{\noindent\makebox[0pt][r]{\@thefnmark\,}#1}
\makeatother 
\renewcommand{\figurename}{\small{Fig.}~}
\sectionfont{\bf\Large}
\subsectionfont{\normalsize}
\subsubsectionfont{\bf}
\setstretch{1.125} 
\setlength{\skip\footins}{0.8cm}
\setlength{\footnotesep}{0.25cm}
\setlength{\jot}{10pt}
\titlespacing*{\section}{0pt}{4pt}{4pt}
\titlespacing*{\subsection}{0pt}{15pt}{1pt}

\fancyfoot{}
\fancyfoot[LO,RE]{\vspace{-7.1pt}\includegraphics[height=9pt]{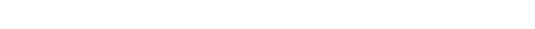}}
\fancyfoot[CO]{\vspace{-7.1pt}\hspace{13.2cm}\includegraphics{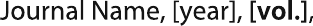}}
\fancyfoot[CE]{\vspace{-7.2pt}\hspace{-14.2cm}\includegraphics{head_foot/RF}}
\fancyfoot[RO]{\footnotesize{\sffamily{1--\pageref{LastPage} ~\textbar  \hspace{2pt}\thepage}}}
\fancyfoot[LE]{\footnotesize{\sffamily{\thepage~\textbar\hspace{3.45cm} 1--\pageref{LastPage}}}}
\fancyhead{}
\renewcommand{\headrulewidth}{0pt} 
\renewcommand{\footrulewidth}{0pt}
\setlength{\arrayrulewidth}{1pt}
\setlength{\columnsep}{6.5mm}
\setlength\bibsep{1pt}

\makeatletter 
\newlength{\figrulesep} 
\setlength{\figrulesep}{0.5\textfloatsep}

\makeatother

\twocolumn[
  \begin{@twocolumnfalse}
\vspace{1em}
\rmfamily
\begin{tabular}{m{2cm} p{13.5cm} }

& \noindent\LARGE{\bf{{The effect of grain shape and material on the nonlocal rheology of dense granular flows$^\dag$}}} \\
\vspace{0.3cm} & \vspace{0.3cm} \\

 & \noindent\large{Farnaz Fazelpour,$^{\ast}$\textit{$^{a}$} Zhu Tang,$^{\ast}$\textit{$^{a}$} and Karen E. Daniels\textit{$^{a}$}} \\
\\
& \noindent\normalsize{Nonlocal rheologies allow for the modeling of granular flows from the creeping to intermediate flow regimes, using a small number of parameters. In this paper, we report on experiments testing how particle properties affect model parameters, using particles of three different shapes (circles, ellipses, and pentagons) and three different materials, including one which allows for measurements of stresses via photoelasticity. Our experiments are performed on a quasi-2D annular shear cell with a rotating inner wall and a fixed outer wall. Each type of particle is found to exhibit flows which are well-fit by nonlocal rheology, with each particle having a distinct triad of the local, nonlocal, and frictional parameters. While the local parameter $b$ is always approximately unity, the nonlocal parameter $A$ depends sensitively on both the particle shape and material. The critical stress ratio $\mu_s$, above which Coulomb failure occurs, varies for particles with the same material but different shape, indicating that geometric friction can dominate over material friction.} \\

\end{tabular}

 \end{@twocolumnfalse} \vspace{0.6cm}

  ]

\renewcommand*\rmdefault{bch}\normalfont\upshape
\rmfamily
\section*{}
\vspace{-1cm}


\footnotetext{\textit{$^{a}$~Department of Physics, North Carolina State University, Raleigh, NC, USA.}}

\footnotetext{\dag~Electronic Supplementary Information (ESI) available. See DOI: }



\section{Introduction}
While idealized studies of granular materials most commonly use circular-shaped particles, these do not correspond to the majority of granular materials present in industrial and geophysical applications. In this paper, we examine the effect of particle properties on the rheology of granular flows, both experimentally and by fitting to a nonlocal model which has previously been validated only for circular particles \cite{Kamrin2012,henann2013predictive,kamrin2015nonlocal,zhang2016,tang2018nonlocal}. Similar to what has been done for local rheological modeling of faster flows  \cite{azema2009influence,azema2018inertial,azema2012discrete,azema2009quasistatic,saint2011rheology}, it is necessary to determine which aspects of the constitutive laws are affected by various particle properties. 

The study of the rheology of granular materials is based on quantifying the relationship between the stress applied to the material, and the resulting flow. We use the dimensionless inertial number $I$ to describe the speed of the flow \cite{forterre2008flows}:
\begin{equation}
I \equiv \frac{\dot{\gamma} d}{\sqrt{P/\rho}}.
\label{eq:I}
\end{equation}
This represents the ratio between a microscopic time $T=d/\sqrt{P/\rho}$ (for particle diameter $d$, particle  material density $\rho$, and the local pressure $P$) and a macroscopic timescale $1/\dot{\gamma}$, which is the mean deformation time under shear rate $\dot{\gamma}$. Large values of  $I$ correspond to rapid flow, while small values are slow, even creeping. In this paper, we focus on 2D experiments in the regime $ 10^{-7} < I < 10^{-4}$, where it is possible to both individually-track particles and to measure the boundary stresses \cite{tang2018nonlocal}; for some particles, we also obtain stress measurement within the bulk, through the use of photoelasticity \cite{daniels_photoelastic_2017,abed_zadeh_enlightening_2019}.
The nondimensional stress ratio is characterized by the ratio between the shear stress $\tau$ and the pressure $P$: 
\begin{equation}
\mu \equiv \frac{\tau}{P}.
\label{eq:mu}
\end{equation}

In local rheology, there is no flow at locations where $\mu$ is less than a  yield criterion $\mu_s$. For slow flows, it has been observed that this criterion fails to explain a number of experimental results \cite{koval2009annular,midi2004dense,Cheng2006,nichol2010flow,reddy2011evidence}. The recent development of nonlocal rheologies \cite{Bouzid2013,bouzid2015non,Kamrin2012,henann2013predictive,zhang2016,dsouza_non-local_2020} aims to provide predictive models which correctly account for the observation of flows where $\mu < \mu_s$. We have previously observed that two of these nonlocal models (the cooperative model by \citet{Kamrin2012} and the gradient model by \citet{Bouzid2013}) are able to provide this predictive power in a 2D granular rheometer \cite{tang2018nonlocal} over a variety of packing fractions and flow rates. In this paper, we directly test the dependence of the nonlocal rheology on particle stiffness and particle shape. 

For our experiments, we compare particles of three different shapes (circles, ellipses, pentagons) and three different elastic moduli. The choice of these three shapes allows us to test for the effects of particle anisotropy (circles vs. ellipses) and particle angularity (circles vs. pentagons). Angular particles are particularly interesting because their contacts are of two types: side-side and side-vertex (vertex-vertex contacts are rare).

\subsection{The cooperative model}

The cooperative model is a nonlocal model \cite{Kamrin2012,henann2013predictive,zhang2016} which has been developed to overcome shortcomings of local rheology by including nonlocal effects in a local Bagnold-like granular flow law. This model has been tested for both experiments and simulations in steady state flows. In the cooperative model, the fluidity is defined as:
\begin{equation}
g \equiv \frac{\dot{\gamma}}{\mu}.
\label{eq:g}
\end{equation}
where the shear ratio $\mu$ is defined in Eq.~\ref{eq:mu}. The cooperative fluidity $g$ has the same units as the shear rate $\dot{\gamma}$ ($s^{-1}$).
 
According to local rheology, the inertial number $I$ and the shear ratio $\mu$ have a linear relationship for $\mu$ larger than the yield criterion $\mu_s$ \cite{da2005rheophysics}. Thus, the local rheology $I(\mu)$ relationship is described using Heaviside function $H$:
\begin {equation}
I(\mu)=\frac{(\mu-\mu_s) \, H(\mu-\mu_s)}{b}.
\label{eq:k1}
\end {equation}
where there is no flow for $\mu < \mu_s$. The parameter $b$ is a constant which models the steepness of the $I(\mu)$ relationship. 

Applying Eq.~\ref{eq:k1} to the fluidity relationship Eq.~\ref{eq:g}, we obtain the local cooperative fluidity $g_\mathrm{loc}$:
\begin {equation}
g_\mathrm{loc}(\mu,P) = \frac{(\mu-\mu_s) \, H(\mu-\mu_s)}{b \mu T}.
\label{eq:k2}
\end {equation}
The cooperative fluidity $g$ is composed of two parts. One is the local rheology contribution, and the other arises from the nonlocal rheology described as a Laplacian term after scaling by a length scale $\xi$: 
\begin {equation}
\nabla^2 g = \frac 1 {\xi ^2}(g-g_\mathrm{loc})
\label{eq:k3}
\end {equation}
The length scale $\xi$ is measured in units of the particle diameter $d$, and takes the form:
\begin {equation}
\frac{\xi}{d} = A\sqrt{\frac{1}{|\mu-\mu_s|}}.
\label{eq:k4}
\end {equation}
 Where $A$ is a constant depending on material properties and indicates the nonlocal effects. The length scale $\xi$ is symmetric around $\mu=\mu_s$, and the system's most sensitive regime is near the yield ratio $\mu_s$ \cite{kamrin2015nonlocal}.


\section{Method}

\subsection{ Apparatus \label{sec:apparatus}}

\begin{figure} 
\centering 
\includegraphics[width=0.9\linewidth]{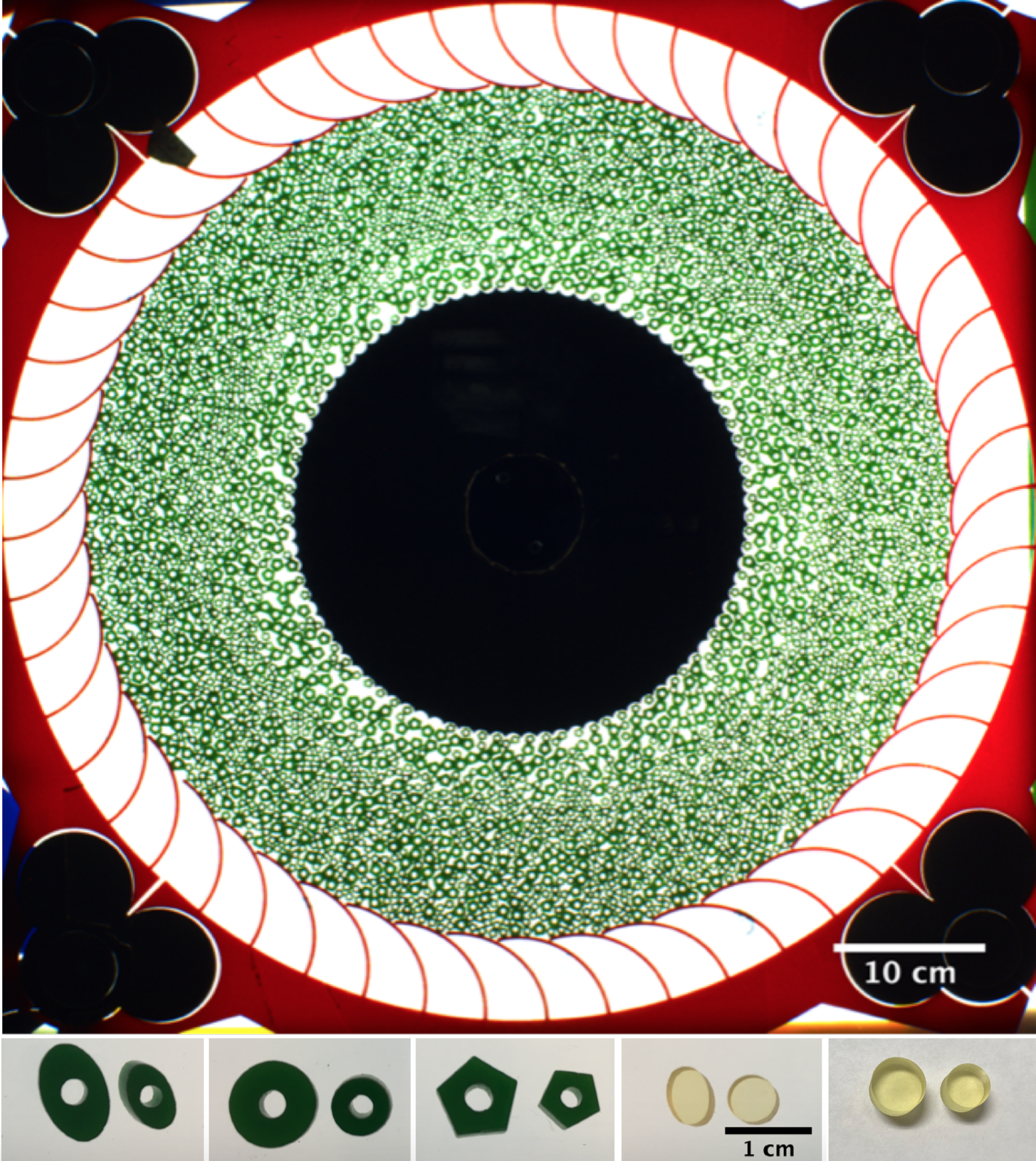} 
\caption{Top: Top view of annular Couette experiment with flat pentagonal particles. Bottom: Photo of the three acrylic shapes: ellipses, circles, and pentagons, along with the circular/elliptical photoelastic particles, and circular photoelastic particles. The central holes allow for easier particle-tracking. Each of the acrylic particles can be circumscribed by a square of side length 0.7 cm (small particles) or 1.0 cm (large particles). 
}
 \label{fig:particle} 
\end{figure}

Our experiments are performed on a quasi-2D annular shear cell with a rotating inner wall and a fixed outer wall. A motor (Parker Compumotor BE231FJ-NLCN with a PV90FB 50:1 gearbox) is attached to the inner wall, providing a constant rotational speed.
We measure the inner wall shear stress $\tau(R_i)$ via a torque sensor (Cooper Instruments \& Systems) attached to the central shaft.
As shown in Fig.~\ref{fig:particle}, the stationary outer wall incorporates 52 laser-cut leaf springs. Each of the springs linearly deforms (both radially and tangentially) under stress from the granular material. Via calibrated image processing \cite{tang_granular_2017}, we obtain quantitative measurements of shear ($\tau$) and normal ($P$) stresses at each of the 52 spring tips. Values are reported as spatial and temporal averages.  
All experiments were performed by rotating the inner wall with speed  $v = 1.1$~cm/s.

We use three types of particles in these experiments. The particles for testing shape-dependence are cut from acrylic (bulk modulus 3 GPa and density $\rho= 1.14$~g/cm$^3$). As shown in Fig.~\ref{fig:particle}, the particles are laser-cut with holes at their centers to aid in particle-tracking. The dimensions of the bidisperse particles, selected to suppress both crystallization and segregation, are as follows. The circles have diameters $d_L=1.0$~cm and $d_S = 0.7$~cm. The ellipses have (minor, major) axes  (0.81~cm, 1.21~cm) and (0.57~cm, 0.85~cm), with $d$ defined as the geometric mean of the major and minor axes , selected to match the same values as the circles. The pentagons have side lengths of 0.65~cm (large) and 0.46~cm (small), so that for $d$ taken to be the distance from one side to the opposite vertex these also match the same values as for the circles. In all cases, the ratio of large to small particle is 1:2 by number, corresponding to approximately equal areas for the two components. 

For testing the effects of material stiffness, we added two additional particle types. First, the particles used in \citet{tang2018nonlocal}: these are a bidisperse mixture of circular (60$\%$) and elliptical (40$\%$) disks cut from $3$~mm thick PhotoStress Plus PS-3 polymer from the Vishay Measurements Group (bulk modulus $0.21$~GPa and density $\rho = 1.15$~g/cm$^3$), these are referred to as Vi3. Second, the particles used in \citet{owens_acoustic_2013,owens_sound_2011}: these are a bidisperse mixture of $d_S = 0.9$~cm and $d_L = 1.1$~cm  circles in equal concentrations, cut from $6.35$~mm thick Vishay PhotoStress material PSM-4 (bulk modulus $E4$~MPa and density $\rho=1.06$~g/cm$^3$), these are referred to as Vi4. Because Vi4 is made of a soft photoelastic material, it allows for the visualization of internal forces, as shown in Fig.~\ref{fig:pressure}(a). By solving an inverse problem on the fringe pattern within each disk, we measure the vector force at each contact, resulting in knowing the shear ($\tau$) and normal ($P$) stresses throughout the material. Details about this process are available in \cite{abed_zadeh_enlightening_2019,kollmer_photo-elastic_nodate,liu_spongelike_2021,daniels_photoelastic_2017,fazelpour_effect_2021}. Unlike for the other particles (acrylic and Vi3) where we measure $\tau$ and $P$ only at the boundaries (torque sensor and leaf spring calibration), these photoelastic particles provide a more quantitative validation of the nonlocal rheology.

\begin{figure}
\centering
\includegraphics[width=0.8\linewidth]{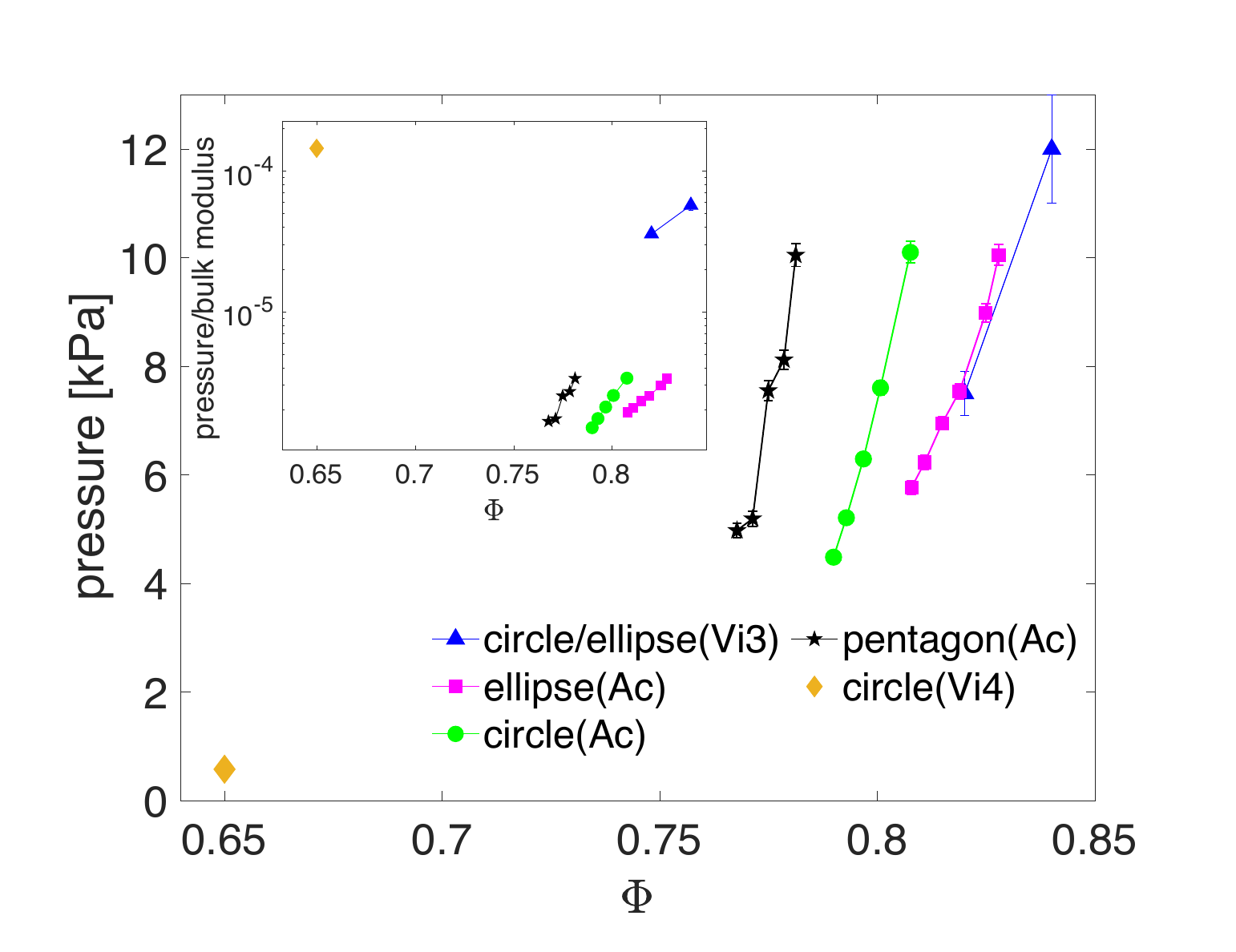}

\caption{Experimentally-determined relationship between the packing fraction $\phi$ and the pressure $P$ for different particle types, {all sheared at an inner wall rotation rate of $v(R_i)=1.1$~cm/s}. In the legend, Vi3 represents particles cut from Vishay PS-3 (data taken from \cite{tang2018nonlocal}), Vi4 represents particles cut from Vishay PSM-4, and Ac represents particles cut from acrylic. 
The error bars represent standard error, with statistics taken over both time and the 52 leaf springs measured for Ac and Vi3 particles, and temporal and spatial standard error for Vi4 particles. The inset is showing the relationship between the ration of pressure to bulk modulus of each particle sets and packing fraction $\phi$.
}
 \label{fig:pfp} 
\end{figure}

Importantly, different particle shapes have dramatically-different packing densities. Random close packing (RCP) for discs  \cite{voivret2007space}, ellipses \cite{delaney2005random}, and pentagons \cite{wang2015structural} are 0.84, 0.895, and 0.80, respectively. In order to conduct experiments at approximately constant pressure, we mapped out the relationship between packing fraction $\phi$ and the measured pressure $P$ for runs at consistent rotation rate. This data is shown in Fig.~\ref{fig:pfp}; for our rheological measurements, we selected a value of $\phi$ to achieve one of two values of pressure: $P = 7.5$~kPa and $P = 10.0$~kPa. Since the  Vi4 particles are several orders of magnitude softer than the other particles, we performed experiments at lower pressure ($P = 0.58$~kPa). In  Fig.~\ref{fig:pfp},  we plot the pressure (or rescaled pressure) as a function of packing fraction.

\begin{table*}
\centering
\footnotesize
 \caption{{\bf Summary of the datasets.} The inner wall rotation $v(R_i)$ is the speed set by the motor controller. 
The number of particles is set by hand to provide the target pressure. The microscopic timescale $T$ is calculated based on measured $P$ and the known values of $d$ and $\rho$. The first column contains data from \citet{tang2018nonlocal}.}

   \begin{tabular}{|c||c|c|c|c|c|c|c|c|}
    \hline    
material &acrylic &acrylic &acrylic &acrylic&acrylic&acrylic &Vishay PS-3 &Vishay PSM-4  \\ \hline
shape  &ellipses &ellipses &circles &circles &pentagons &pentagons&circles/ellipses & circles  \\ \hline
    $v(R_i)$ [$d$/s]  &1.3 &1.3 &1.3 &1.3&1.3 &1.3 &2 &1.1  \\ \hline
    \# of particles       &3242 &3210 &2920 &2895  &3438 &3426 &5610 &1724 \\ \hline
$P$ [kPa] & 10$\pm$0.2  & 7.5$\pm$0.1& 10$\pm$0.2& 7.5$\pm$0.1&  10$\pm$0.2 & 7.5$\pm$0.1 & 7.5$\pm$0.4 & 0.58$\pm$0.02\\ \hline    
$S_0$ [kPa] & 3.5 $\pm$0.6 & 1.4$\pm$0.3 &2.9 $\pm$0.5 & 1.6$\pm$0.3& 2.2 $\pm$0.4& 1.5 $\pm$0.3 & 0.8$\pm$0.2 & 0.18$\pm$0.04 \\ \hline    

$T$ [msec] & 2.9$\pm$0.1 & 3.3$\pm$0.1& 2.9 $\pm$0.1& 3.3 $\pm$0.1&  2.9 $\pm$0.1 & 3.3 $\pm$0.1 & 2.2$\pm$0.1 & 13.5$\pm$0.1 \\ \hline    
  \end{tabular}
\label{tbl:pressure2}
\end{table*}

A summary of all experimental runs is provided in Table.~\ref{tbl:pressure2}, including measurements of the shear stress at the inner wall. Note that we  observe that runs with a higher packing fraction $\phi$ (also higher pressure $P$) have a higher inner wall shear stress $S_0$. Each experiment's microscopic timescale $T$ is calculated from the associated pressure measurement $P$, together with particle properties; all are approximately 3 msec, except for Vi4 which has a timescale several times larger due to the lower pressure. 

\subsection{Measuring speed and shear rate \label{sec:speed}}

\begin{figure} [h]
\centering 
\includegraphics[width=0.8\linewidth]{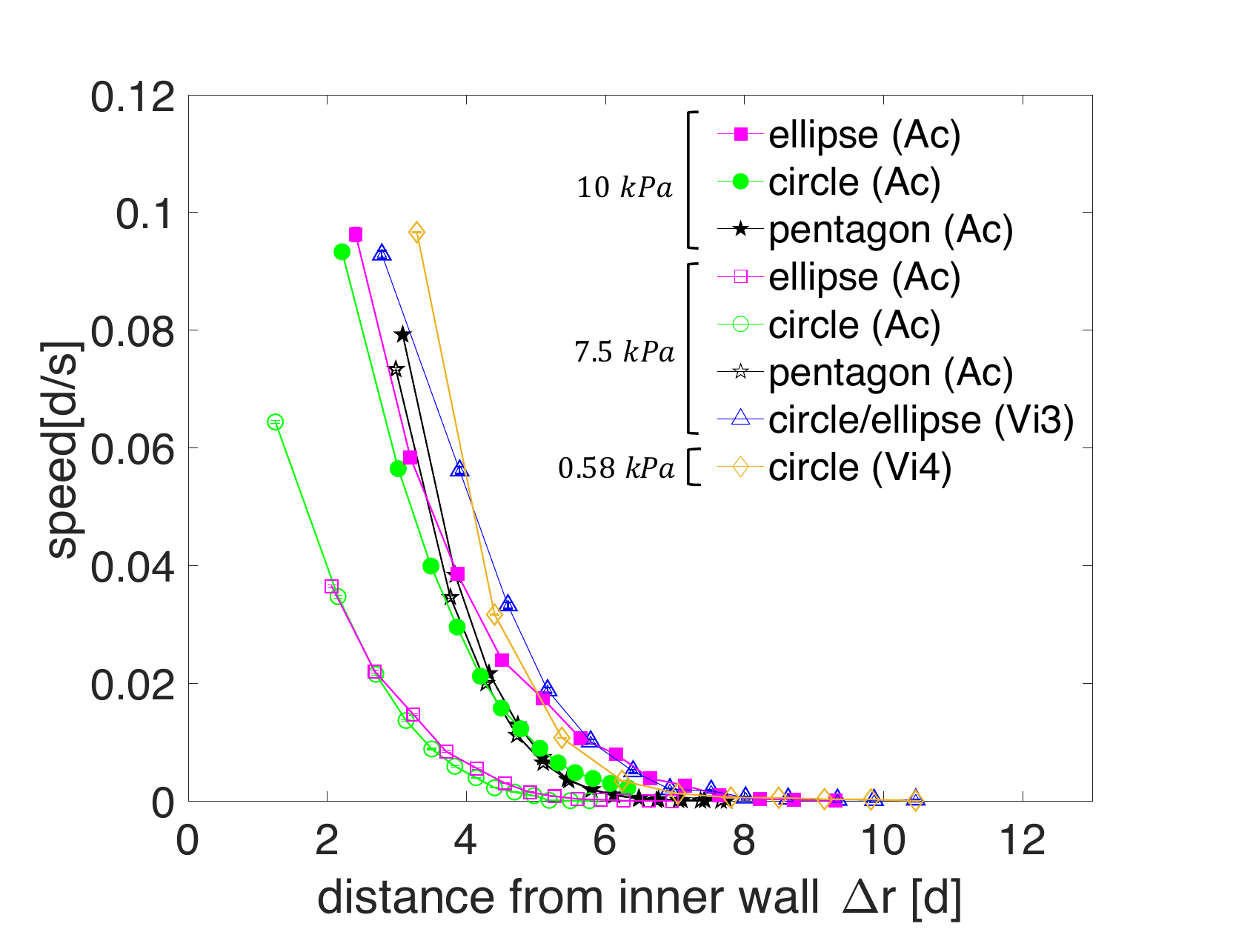} 
\includegraphics[width=0.8\linewidth]{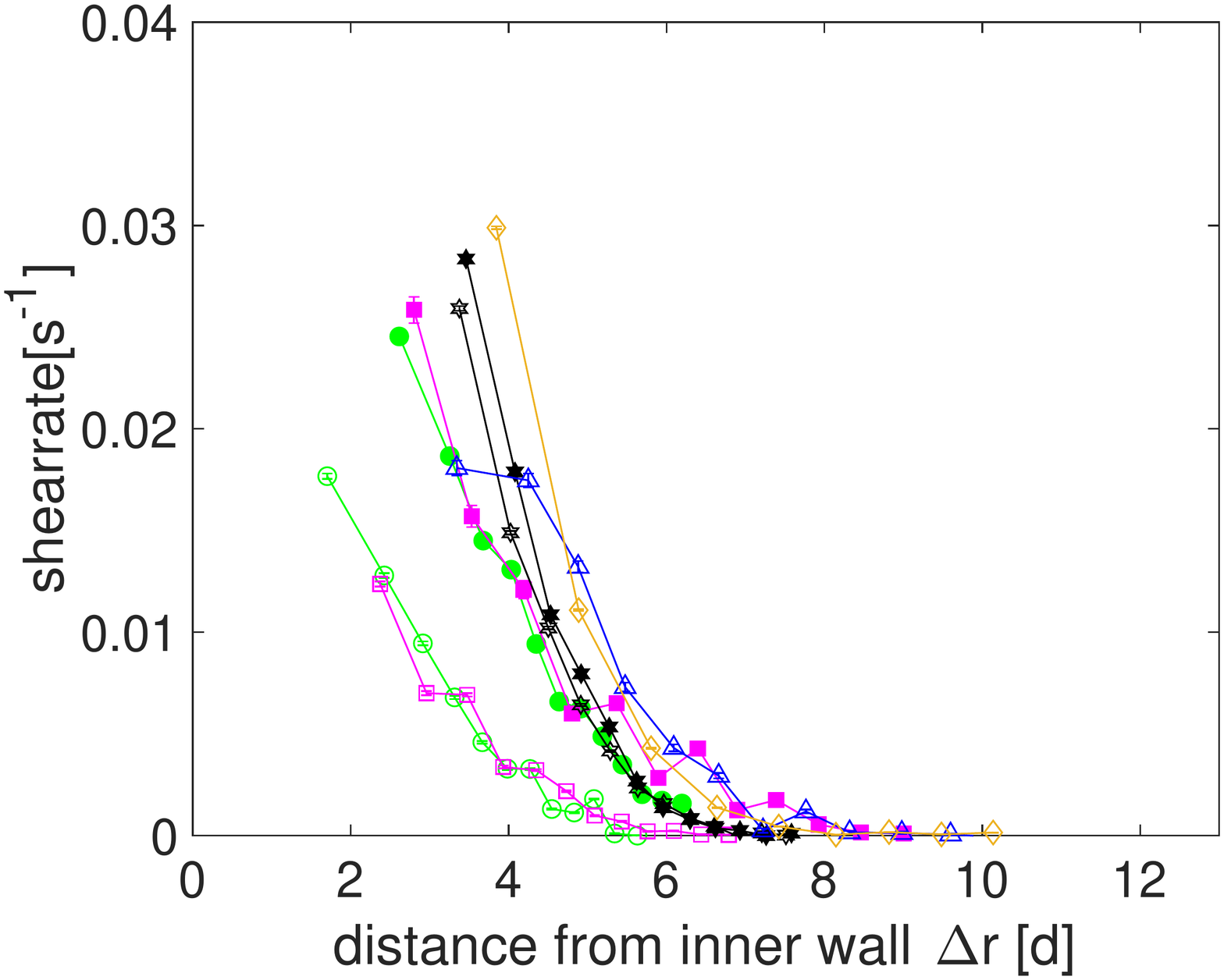} 
\caption{Measured  (a) tangential speed profile $v(r)$ and (b) shear rate $\dot{\gamma}(r)$. Blue triangle data is reproduced from \citet{tang2018nonlocal}. 
}
 \label{fig:spd} 
\end{figure}

We locate the centroids of the particles using Matlab's Hough transform \cite{HT}, and create space-time trajectories using the  Blair-Dufresne particle-tracking algorithm \cite{DB}. The tangential speed profile $v(r)$ is calculated in within concentric rings of width 0.5~$d$. To calculate the shear rate, we use Fourier-derivatives as described in  \citet{tang2018nonlocal}. Note that, due to the annular geometry, the shear rate is given by ${\dot{\gamma}}(r) = \frac{1}{2} \left( \frac {\partial v}{\partial r}- \frac{v}{r} \right)$. These measurements are presented in Fig.~\ref{fig:spd}, analyzed from $10^4$ frames analyzed for each dataset. Observe that for circular and elliptical particles, the runs at lower $P$ have smaller values of $v$ and $\dot{\gamma}$ when compared at the same distance from the inner wall. Pentagonal particles do not exhibit this dependence.

\subsection{Estimating basal friction effects\label{sec:cg}}

\begin{figure} 
\centering 
\includegraphics[width=0.8\linewidth]{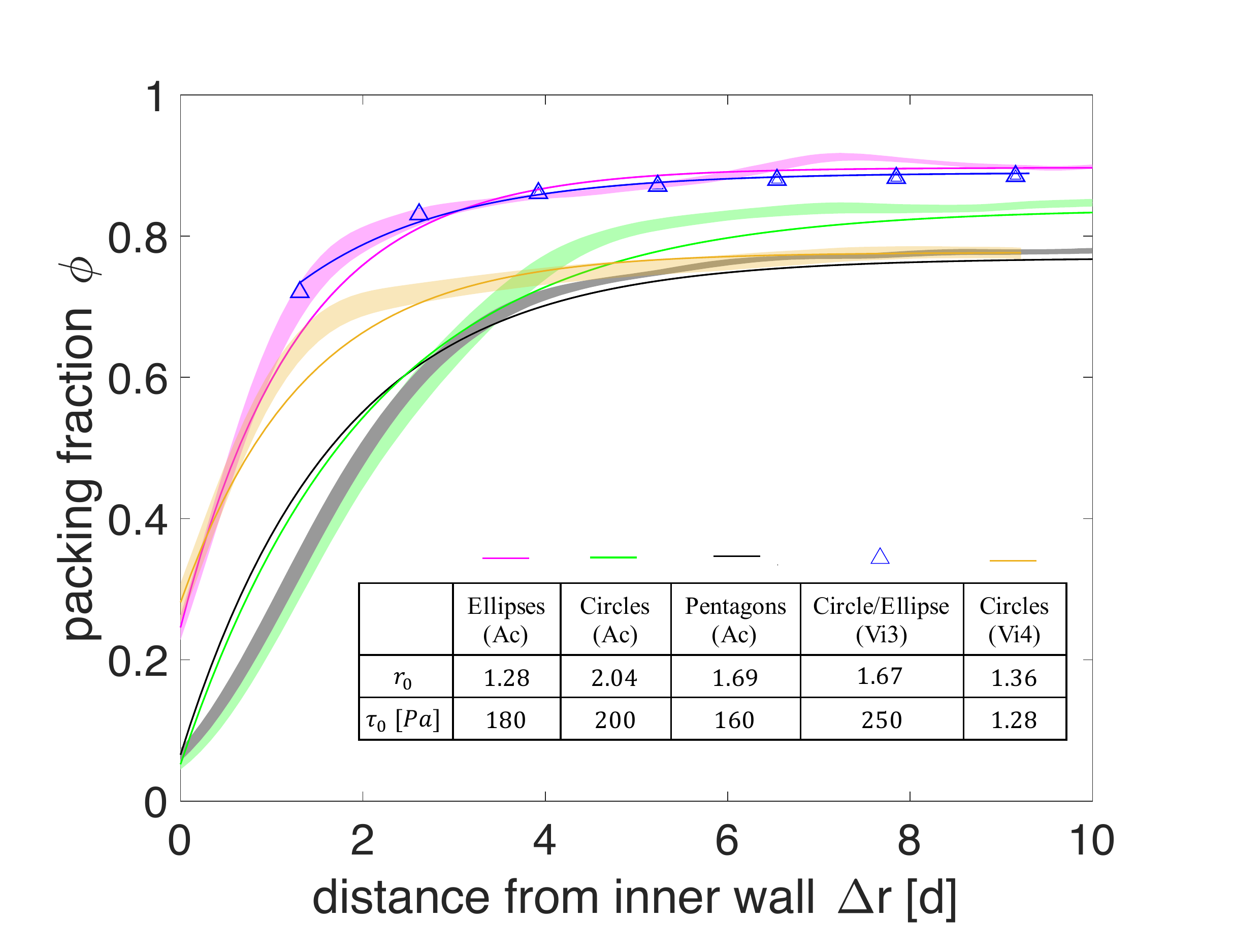} 
\caption{The local packing fraction for different particle shapes and materials at $P=7.5$kPa ($P=0.58$kPa for Vi4 particles) is calculated by coarse-graining method. The solid curves are the corresponding fitting curves from Eq.~\ref{eq:pf}. The error bars are the standard error. The circle/ellipse (Vi3) data are from \citet{tang2018nonlocal}. The inset table shows fitting parameters in Eq.~\ref{eq:pf} and Eq.~\ref{eq:tor2} for each particle shape and material. These parameters are the same for the same particles at different pressure.
}
 \label{fig:particlep} 
\end{figure}

As previously reported in \citet{tang2018nonlocal}, it is necessary to account for the basal friction in order to correctly measure the local shear stress on each particle. Therefore, we again assume that the stress from basal friction is proportional to the local packing fraction $\phi$. We calculate $\phi(r)$ for each particle type using the coarse-graining method of \citet{weinhart2013coarse}, based on the tracked locations of all particles. For the Lucy function used for coarse-graining, we find that a width parameter of $w=1.3$ is sufficient to remove major fluctuations without being over-smoothed.  
As shown in Fig.~\ref{fig:particlep}, the data can be approximated by an exponential with a decay parameter $r_0$. We fit each curve to the equation
\begin{equation}
\phi(r)=\phi_0 \left[ 1-e^{-\Delta r/r_0} \right]+ \phi(R_i).
\label{eq:pf}
\end{equation}
where the parameter $\phi_0 \equiv \phi(R_o)-\phi(R_i) $ is the difference in packing fraction between the outer wall and the inner wall. We observe that mixtures containing elliptical particles pack the most densely, followed by circular particles and then pentagons, as also observed in Fig.~\ref{fig:pfp}.

Using Eq.~\ref{eq:pf}, we again write a phenomenological model for the shear stress: 
\begin {equation}
 \tau(r)= S_0 \left( \frac{R_i}{r} \right)^2 + \tau_0 \left [1-e^{-\Delta r/r_0} \right]
 \label{eq:tor2}
\end{equation}
This model consists of two parts. The first term $S_0 \left( \frac{R_i}{r} \right)^2$ arises from the annular geometry, with the parameter $S_0$ corresponding to the torque measured at the central shaft. The second term arises from the basal friction, with the parameter $r_0$ taken from the fits to the packing fraction curves  shown in Fig.~\ref{fig:particlep}. 

\begin{figure} 
\centering 
\includegraphics[width=\linewidth]{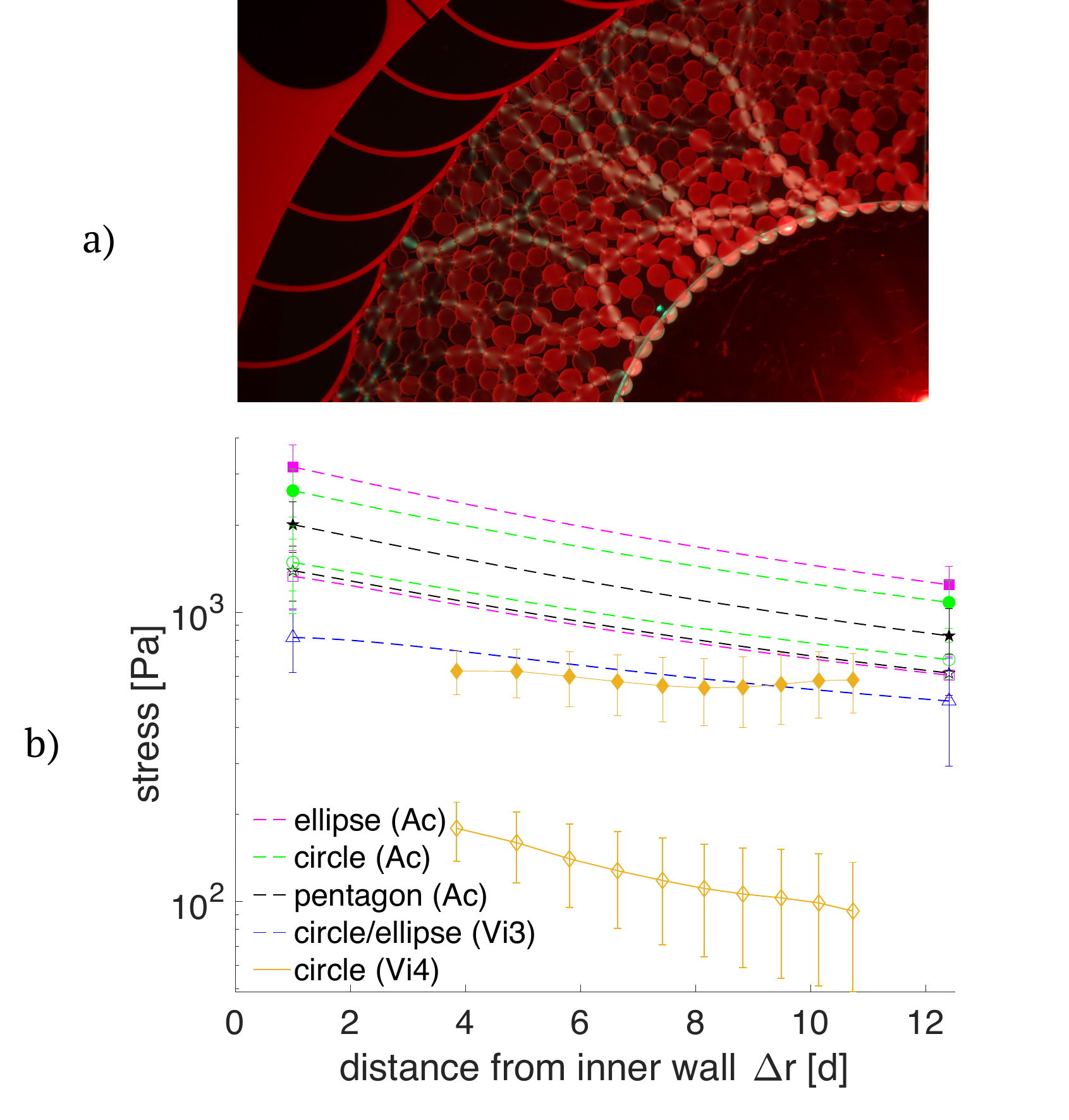} 
\caption{(a) Sample image Vi4 particles under shearing. The green patterns indicate internal forces. See the Supplementary Information for a video of force chains evolution under white light. (b)Shear stress measured from boundaries for Ac and Vi3 particles shown by filled-symbols (at pressure $10$kPa) and empty-symbols (at pressure $7.5$kPa). The dashed-lines are full shear stress profiles for Ac and Vi3 particles obtained from measured shear stress at the boundaries and from Eq.~\ref{eq:tor2}.The same shape is in the same color. The full stress profiles are measured by photoelasticity for Vi4 particles . The shear stress shown by empty-diamond and the pressure shown by solid-diamond. 
}
 \label{fig:pressure} 
\end{figure}

For the acrylic and Vi3 particles, the parameter $\tau_0$ is calculated from $\tau(R_o)-S_0 \left(\frac{R_i}{r} \right)^2 $ with $\tau(R_o)$ measured from the leaf springs. The full shear stress profiles are obtained using Eq.~\ref{eq:tor2} (dashed-lines Fig.~\ref{fig:pressure}(b)) and boundaries (symbols Fig.~\ref{fig:pressure}(b)). 
For both the high and low pressure datasets taken for the same particle shape, we use the same set of $(r_0, \tau_0)$ parameters since the difference in particle number is only about $1\%$.  These values are given in the inset to Fig.~\ref{fig:particlep}. 

For the Vi4 particles, we perform photoelastic stress measurements on $2000$ frames (taken at $0.2$ Hz, see sample image in Fig.~\ref{fig:pressure}(a)) and time-average the  coarse-grained stress field calculated from the vector contact forces. This provides the shear stress $\tau$ and pressure $P$ throughout the material, as shown in Fig.~\ref{fig:pressure}(b). For this dataset, the parameters in Eq.~\ref{eq:tor2} can be found with $S_0$ and $\tau_0$ obtained directly from fitting shear stress profiles, and $r_0$ obtained from packing fraction fitting (Fig.~\ref{fig:particlep}).

These different methods of measurements have advantages and disadvantages. For instance, the boundary stress measurements (for Ac and Vi3 particles) gives stress values directly at the inner and outer wall, but we cannot measure stress directly all throughout material. The  photoelasticity method (Vi4) measures stress throughout the material, but falls short of the  inner and outer walls due to lighting imperfections and the coarse-graining  length scale.




\section{Results \label{sec:compare}}

\begin{figure} 
\centering
\includegraphics[width=0.8\linewidth]{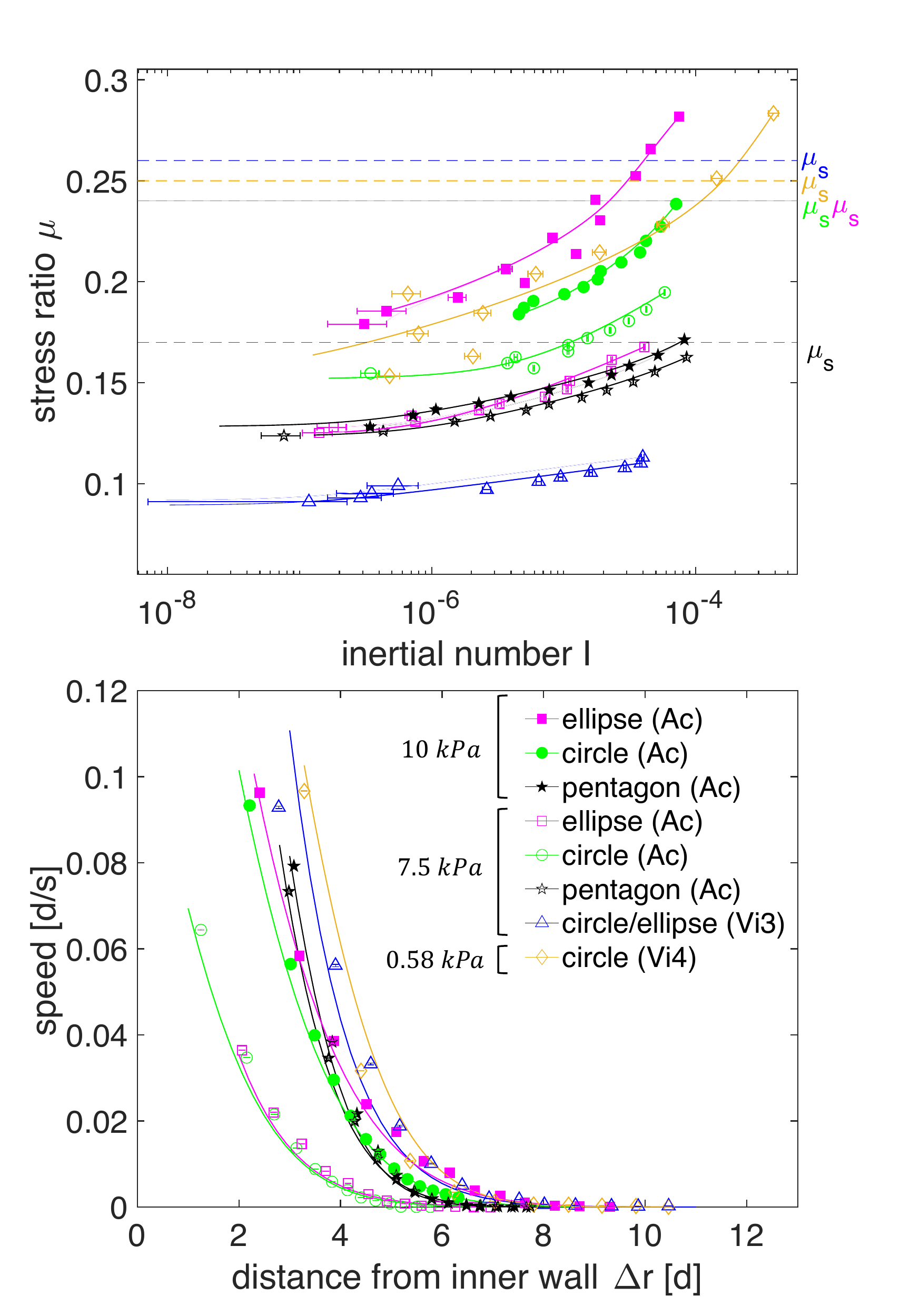}
\caption{(a) Stress ratio $\mu$ as a function of inertial number $I$, for all eight datasets. (b) Speed profiles $v(r)$ for all eight datasets. The solid curves are calculated from the cooperative model, using model parameters from Table~\ref{tbl:parameters2}.
}
 \label{fig:bestfit} 
\end{figure}

The comparison that follows utilizes the cooperative model of \citet{Kamrin2012}, using methods previously described in \citet{tang2018nonlocal}. 
In Fig.~\ref{fig:bestfit}a, we plot the experimentally-measured  $\mu(I)$ relationships for all three particle materials and all three particle shapes. Note that in all cases, the low-$P$ run lies at lower $\mu$ than the high-$P$ run for the same particles shape, which arises because the shear stress $\tau$ decreases even faster than $P$. For example, for the runs using elliptical particles, the inner wall stress $\tau(R_i)$ drops from 3500 Pa to 1400 Pa, for only a 25\% decrease in pressure. 

In all cases, we are able to fit the experimentally-determined $\mu(I)$ data using the cooperative model, using the parameters listed in Table~\ref{tbl:parameters2}. These were determined as follows.

\subsection{Rheological parameters \label{sec:params}}

\begin{table}
\footnotesize
\centering
\caption{Nonlocal parameters $A$ and $b$ for different particle shapes/materials. These parameters are the same for the same particles at different pressures.}
    \begin{tabular}{|c|c|c|c|c|c|}
  \hline
       ~    & acrylic       & acrylic        & acrylic   &  Vishay PS-3 &  Vishay PSM-4    \\ 
       ~       & ellipses       & circles        & pentagons  & circles/ellipses &circles \\ \hline
     $\mu_s$ & $0.24\pm0.02$ & $0.24\pm0.02$ & $0.17\pm0.01$ & $0.26\pm0.01$ & $0.25\pm0.01$ \\ \hline
      $b$ & $1.1\pm0.5$& $1.1\pm0.5$ & $1.1\pm0.5$ & $1.1\pm0.5$ & $1.1\pm0.5$ \\ \hline
      $A$  & $0.28\pm0.01$ & $0.30\pm0.02$ & $0.13\pm0.02$ & $0.41\pm0.02$ & $0.17\pm0.03$ \\ \hline
      \end{tabular}
  \label{tbl:parameters2}
\end{table}

For the six datasets taken for acrylic and Vi3 particles, we performed the same particle-tracking and boundary stress measurements as done in \citet{tang2018nonlocal}. For the dataset taken for Vi4 particles the same particle-tracking is performed, but the stress measurements are done using photoelastic measurements (see Fig.~\ref{fig:pressure}). To measure the yield stress ratio $\mu_s$, we performed an additional run (not shown in Table~\ref{tbl:pressure2}) at $v(R_i)=0.0013d$/s, for which we previously observed that the ratio of the inner wall shear stress to the pressure is a good estimate \cite{tang2018nonlocal}. These values are shown in Table~\ref{tbl:parameters2}. We observed that all circular/elliptical particles have a similar value of $\mu_s$, while the angular particles yield at a much lower stress ratio. This reflects that the shape of the interparticle contacts (rounded vs. angular) is an important control on $\mu_s$, beyond material properties such as coefficient of friction or elastic modulus \cite{papanikolaou2013isostaticity}. 

For each dataset, we find the fluidity profile $g(r)$ by solving Eq.~\ref{eq:k3} using the Matlab ODE solver. We set the boundary conditions empirically, by measuring  $g$ near the wall. To obtain values for parameters $(A,b)$, we use Levenberg-Marquardt optimization to fit the dataset for each particle material, and for each particle shape, we obtain $(A,b)$ at $P = 10$~kPa , and then apply these parameters to the dataset collected at $P = 7.5$~kPa.  The resulting parameters are shown in Table~\ref{tbl:parameters2}, and the best fitting curves are shown in Fig.~\ref{fig:bestfit}. 
Values of the nonlocal parameter $A$ are similar (but not identical) for rounded particles with the same material, but differ for pentagonal particles. Particle material plays important role in the values of the nonlocal parameter $A$, rounded particles with different material have different nonlocal parameter $A$. Values of the local parameter $b$ are insensitive to the particle properties, whether material or shape.


\subsection{Length scale \label{sec:length}}

\begin{figure} [t]
\centering
\includegraphics[width=0.8\linewidth]{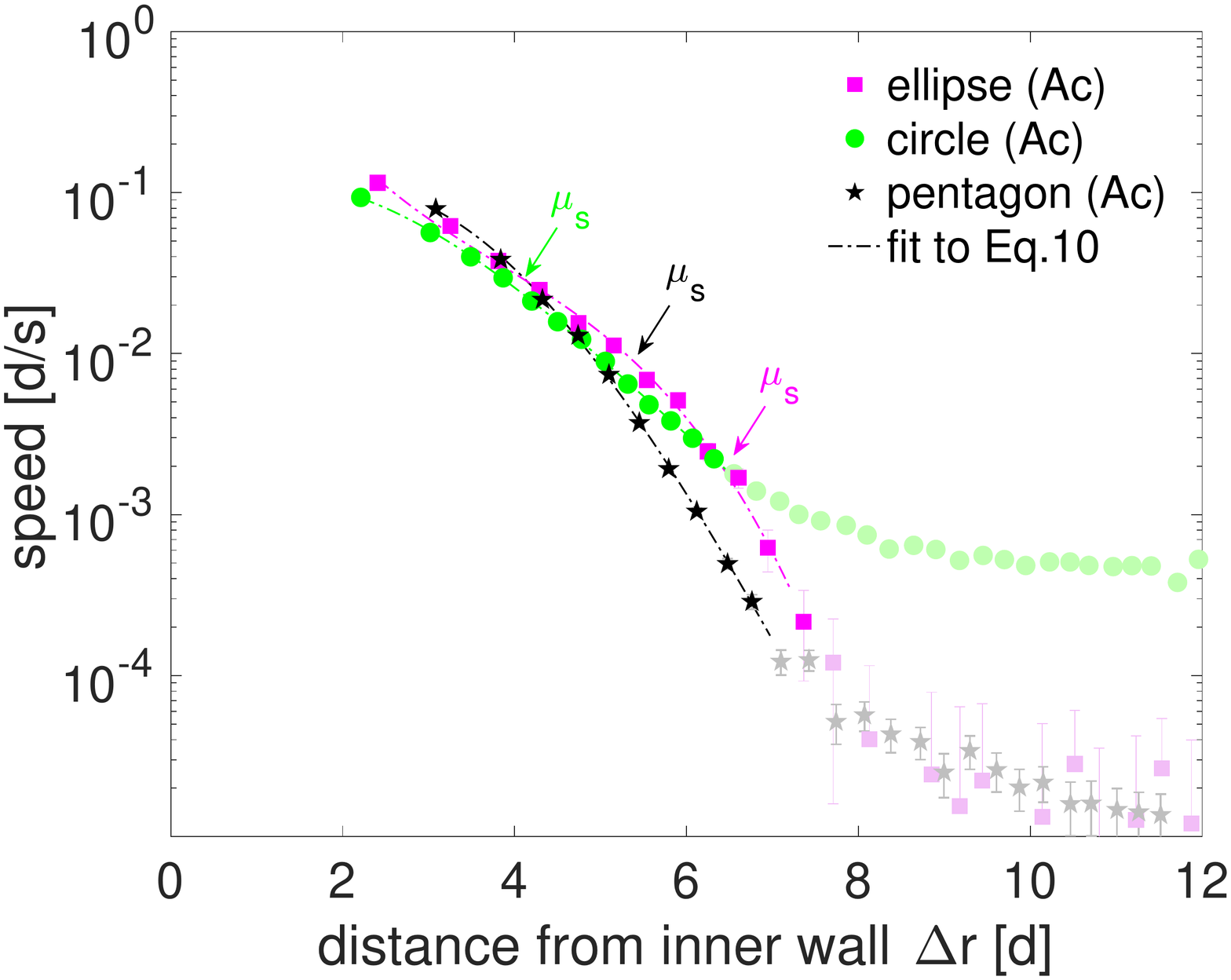}
\includegraphics[width=0.8\linewidth]{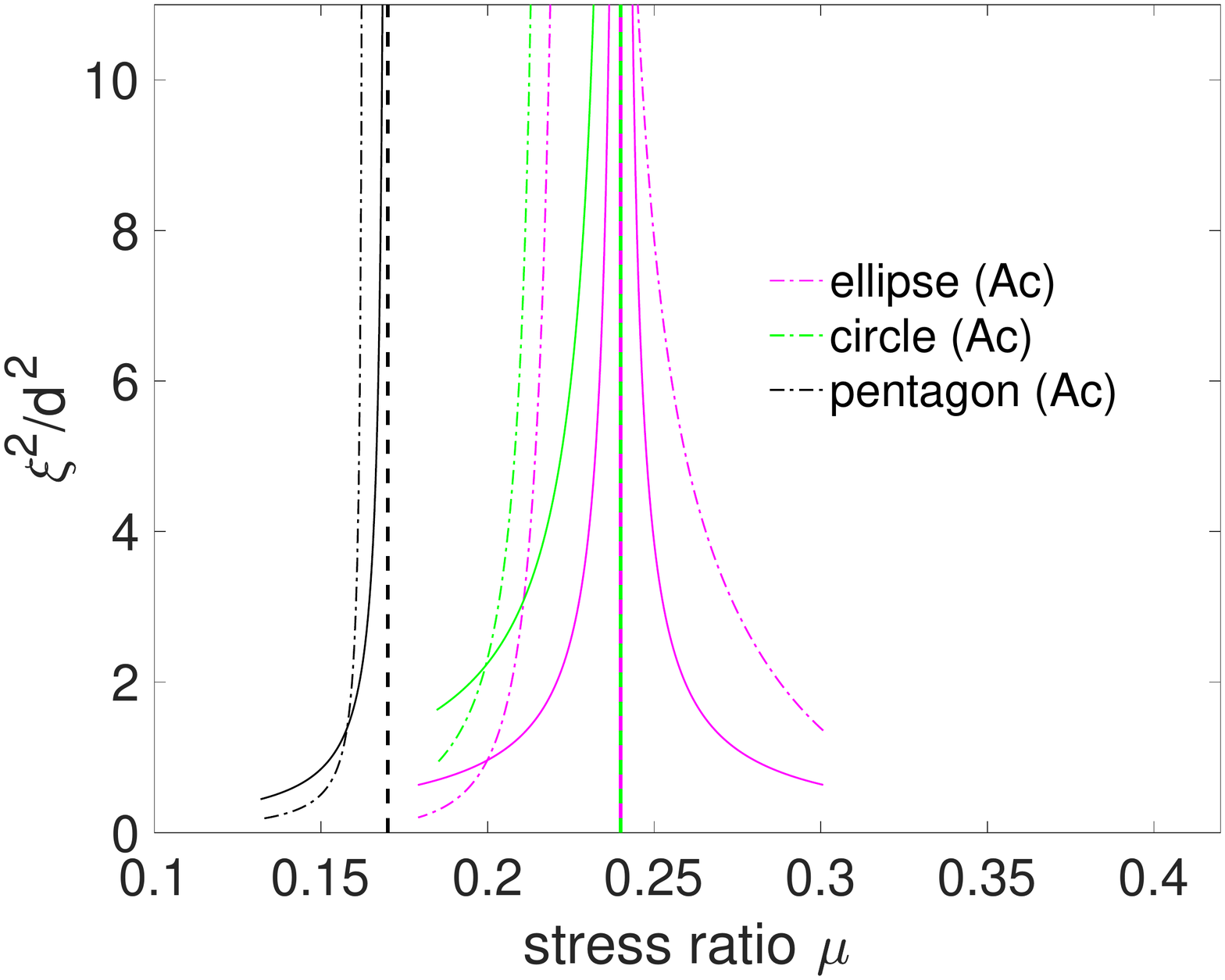}
\caption{(a) Speed profile for $P = 10$~kPa particles in different shapes. The dash-dotted lines are fitted by Eq.~\ref{eq:fit_speed} Brighter data points are fitting data, lighter data points are not involved in the fitting of speed curves. (b) Comparison of measured length scale (the dash-dotted curves) to the theoretical curves (solid curves). The same shape is in the same color.
}
 \label{fig:xi} 
\end{figure}

The lengthscale $\xi$ in Eq.~\ref{eq:k4} represents the influence of the nonlocal term in the vicinity of $\mu_s$. To check this dependence and the effects of particle shape on it independently, we pick three different particle shapes all cut from acrylic (at pressure $P = 10$~kPa), to determine whether the assumptions of the model are justified. The measured lengthscale is calculated by Eq.~\ref{eq:k3} using the analytical method presented in \citet{tang2018nonlocal}, and the theoretical curves are calculated by Eq.~\ref{eq:k4}.

To perform this validation, we perform an empirical fit to the speed profile $v(r)$, in order to take the necessary higher-order derivatives. We (as before)  observe that the empirical speed profile function is well-fit by Eq.~ \ref{eq:fit_speed}.
\begin{equation}
v(r)  = v_0 \mathrm{exp}\left[ \alpha_3 r^3+ \alpha_2 r^2+ \alpha_1 r+ \alpha_0 \right]
 \label{eq:fit_speed}
\end{equation}

The resulting fits in the vicinity of $\mu_s$ are shown in Fig.~\ref{fig:xi}. Note that $v(r)$ for the circular particles fails to fall off as quickly as was observed for the elliptical and pentagonal particles. In the original movies for these experiments, we  can confirm this observation, and additionally observe that this run exhibits significant crystallization effects. Because crystallized domains are more stable under higher pressure than lower pressure \cite{khain2006shear}, they are more efficient at transmitting shear at larger distances from the shearing surface. Since the calculation of $\xi$ does not take place in this outer region, we are able to proceed with the model validation. 

From  Eq.~\ref{eq:k4}, we expect a divergence of $\xi(\mu)$ at $\mu_s$, and the kinematics of the particles thereby provides an independent measurement of $\mu_s$. We estimate its location by drawing an arbitrary horizontal line in Fig.~\ref{fig:xi}(b), and determining the two intersection points of this horizontal line and the $\xi^2(\mu)$ curve. The measured value of  $\mu_s$  lies at the mean $\xi$ of these two intersections: for circular particles $\mu_s=0.23$, for elliptical particles, $\mu_s=0.22$, and for pentagonal particles, $\mu_s = 0.16$. These values correspond closely to the values measured by quasi-static shearing, given in Table~\ref{tbl:pressure2}. Moreover, we can see variations in the nonlocal effects for different particle shapes in Fig.~\ref{fig:xi}(b) where $\xi^2(\mu)$ takes a wider form in circular/elliptical particles than pentagons.

\section{Microscopic description of granular fluidity \label{sec:phif}}

According to a recent interpretation\cite{zhang2016} of the granular fluidity $g$, the only variables affecting the granular fluidity $g$ are the velocity fluctuation $\delta v$ (the square root of the granular temperature) and the packing fraction $\phi$. This motivates writing the granular fluidity $g$ in a microscopic form:
\begin{equation}
g=\frac{\delta v}{d} F(\phi).
\label{eq:g_micro}
\end{equation}
While the physical origins of the function $F(\phi)$ remains unknown, simulation results from Zhang and Kamrin \cite{zhang2016} showed that the function $F(\phi)$ is independent of the configurations (both the driving speed and the geometry) and only on the packing fraction $\phi$, and may also depend on the particle properties.

\begin{figure} 
\centering
\includegraphics[width=0.8\linewidth]{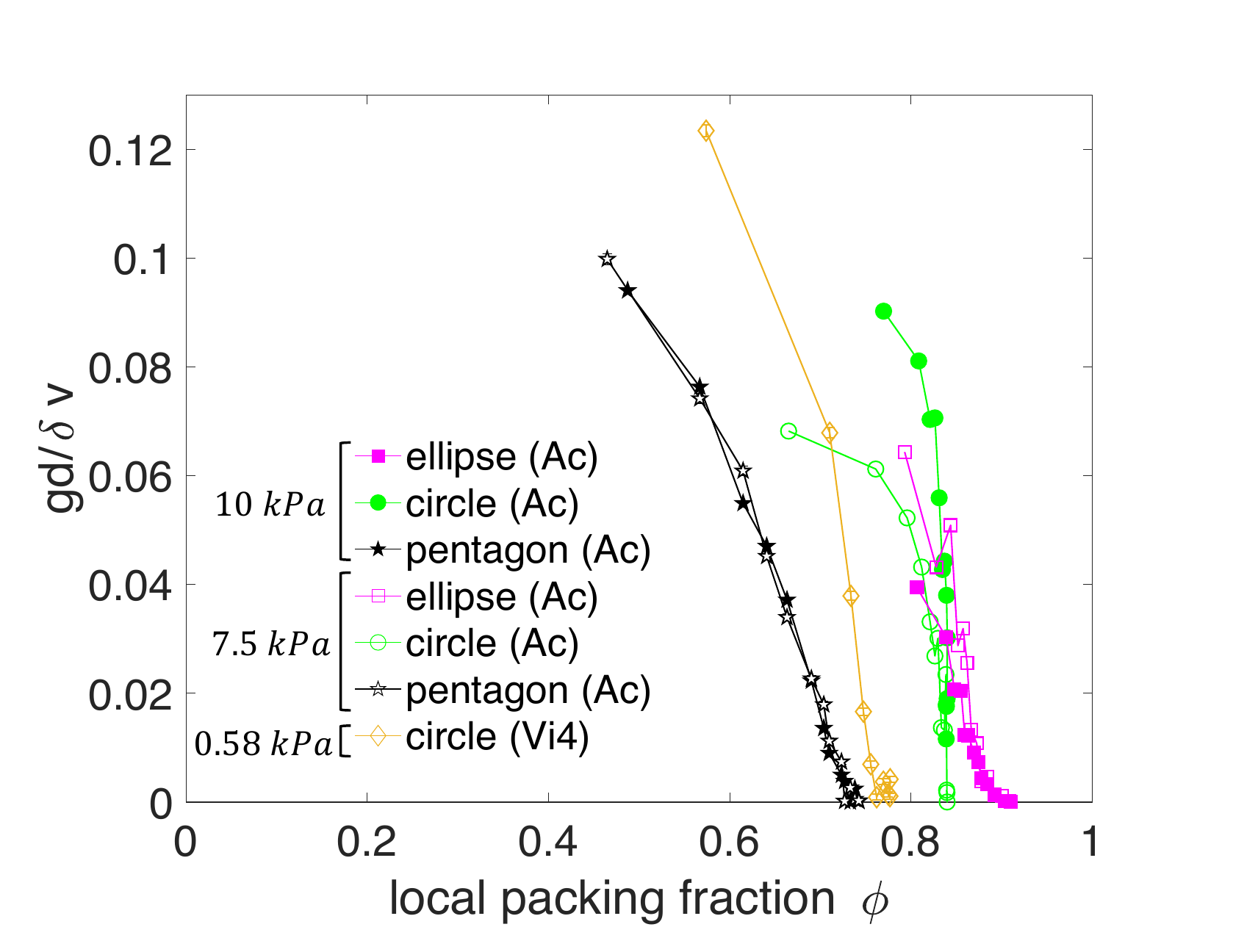}
\caption{Determination of the function $F(\phi)$ from Eq.~\ref{eq:g_micro},  for different acrylic shapes under both pressures and Vi4. Data are from the same bins as used in the $\mu(I)$ plot shown in Fig.~\ref{fig:bestfit}a. The error bars are from the standard error of granular fluidity $g$.}
 \label{fig:fdv} 
\end{figure}

We test this definition using particles of different shape but the same material (acrylic) and same shape(circles) but different material. Here,  we exclude the Vi3 particles since the mixture of ellipses and circles does not have a clear comparison in the other datasets. 
The velocity fluctuations are measured within concentric rings of width $0.5d$, using the same data presented in Fig.~\ref{fig:particlep}. As shown in Fig.~\ref{fig:fdv}, the shape of the particles affects the function $F(\phi)$: each shape and material has its own characteristic curve that ends at $\phi_\mathrm{RCP}$ for that specific shape/material. While the elliptical and pentagonal particles display a consistent shape, independent of pressure, the circular particles do not. Instead, the shape of $F(\phi)$ changes, likely due to the crystallization problems presented earlier. 

\section{Conclusions}

We have established the success of the cooperative nonlocal model \cite{Kamrin2012,kamrin2015nonlocal,zhang2016,henann2013predictive} in describing the rheology of non-circular particles. The particular shape of the particles plays an important role in the particular choice of modeling parameters. While the local parameter $b$ is independent of the particle shape or material, the critical stress ratio $\mu_s$ only depends on the particle shape, and the nonlocal parameter $A$ is strongly sensitive to both the particle shape and stiffness. Nonlocal effects are observed to be more important for rounded particles than for angular particles, as measured by the magnitude of the $A$. For similar shapes, we observe that softer particles have a very different nonlocal effects depending on the softness of the particles. To obtain these parameters $(A,b)$, we set the boundary conditions empirically and see that the results are sensitive to the choice of boundary condition. The open question is how we can set the boundary conditions and what role the walls play in these results.  Finally, we find that the particle-scale definition of granular fluidity takes a similar shape to that observed in simulations \cite{zhang2016}, and may be strongly affected by crystallization.

\section*{Conflicts of interest}
There are no conflicts to declare.

\section*{Acknowledgements}
We thank Michael Shearer and Theodore Brzinski for useful discussions about the project, and Austin Reid for inspiring the boundary wall designs.  We are grateful to the National Science Foundation (NSF DMR-1206808 and DMS-1517291) for the construction of the particles and apparatus, and the International Fine Particle Research Institute (IFPRI) for financial support.


\balance

\bibliography{rbb} 
\bibliographystyle{rsc} 

\end{document}